\documentclass[12pt]{article}
\usepackage{amsmath}
\usepackage{amssymb}
\usepackage[mathscr]{eucal}
\usepackage{authblk}
\usepackage[nocompress]{cite}
\usepackage{enumitem}
\usepackage{pdfsync}
\usepackage[
bookmarks=true,
backref=true,
colorlinks=true,
linkcolor=red,
citecolor=blue,
urlcolor=green]{hyperref}

\newcommand{\gen}[1]{\partial_{#1}}
\newcommand{\pr}[1]{\rm pr^{(#1)}}
\newcommand{\curl}[1]{ \left\{#1\right\} }

\newtheorem{example}{Example}[section]

\numberwithin{equation}{section}
\numberwithin{thm}{section}
\numberwithin{lemma}{section}
\numberwithin{prop}{section}
\numberwithin{cor}{section}
\numberwithin{rmk}{section}
\numberwithin{defn}{section}
\DeclareMathOperator{\Sl}{sl}

\DeclareMathOperator{\cosec}{cosec}

\DeclareMathOperator{\SL}{SL}

\DeclareMathOperator{\Div}{Div}

\renewcommand{\Im}{\operatorname{Im}}

\newcommand{\lie}{\mathfrak g}

\newcommand{\lr}[1]{\langle{#1}\rangle}

\begin{document}

\title{A unified approach to  the time-dependent quantum harmonic oscillator via point transformations and Lie symmetries}

\author[1]{F. G\"ung\"or\thanks{e-mail: gungorf@itu.edu.tr}}
\author[2]{J. Negro \thanks{e-mail: jnegro@fta.uva.es}}

\affil[1]{Department of Mathematics, Faculty of Science and Letters, Istanbul Technical University, 34469 Istanbul, Türkiye}

\affil[2]{Departamento de Física Teórica, Atómica y Óptica, Universidad de Valladolid, 47011 Valladolid, Spain}

\date{}

\maketitle

\abstract{We revisit the time-dependent  harmonic oscillator (TDHO) in quantum mechanics based on the notion of equivalence transformations and Lie point symmetries. Within this framework, we give the most general point transformations depending on seven parameters which map  from the TDHO to the usual harmonic oscillator (HO) and the other way around. As a byproduct we reconstruct the Schrödinger propagator (Mehler's kernel) for the time-independent harmonic oscillator. We demonstrate, without recourse to the Lewis--Riesenwert (LR) invariant, how Lie algebraic methods enable alternative  derivation of the normalized wave functions of the TDHO by constructing  solutions invariant under some specific subgroup of  $\SL(2,\mathbb{R})$ combined with the gauge group.  We further show that, on the relevant $\SL(2,\mathbb{R})$-invariant solution family, the Lewis--Riesenfeld
operator defines a compatible second-order differential constraint, thereby clarifying its relation to the Lie symmetry structure of the TDHO.
}

\section{Introduction}
In a seminal work \cite{Lewis1968}, Lewis made use of Kruskal's asymptotic theory to show the existence of a class of  exact invariants for the time-dependent harmonic oscillator with  Hamiltonian
\begin{equation}\label{hamilt}
  H(t)=\frac{1}{2}[p^2+\Omega^2(t) q^2],
\end{equation}
$q$ is a canonical coordinate, $p=-i\hbar \gen x$ is its conjugate
momentum, $\Omega(t)$ is an arbitrary function. In the quantum case, $q$ and $p$ satisfy the usual
commutation relation $[q,p]=i \hbar$. The so-called Lewis invariant is given by
\begin{equation}\label{Lewis-inv}
  I(t)=\frac{1}{2}[\rho^{-2}q^2+(\rho p-\dot{\rho}q)^2],
\end{equation}
where $\rho$ is a function of $t$ satisfying the Ermakov--Pinney equation
\begin{equation}\label{EP-rho}
  \ddot{\rho}+\Omega^2(t) \rho= \rho^{-3}.
\end{equation}
This invariant is also true in  classical mechanics with the equation of motions are
\begin{equation}\label{classical}
  \ddot{q}+\Omega(t)^2 q=0.
\end{equation}
We note that this invariant can be extended to more general equations of motion using the theory of differential invariants \cite{GuengoerTorres2017}.

In a sequel paper \cite{LewisRiesenfeld1969}, Lewis and Riesenfeld established that $I(t)$ given by \eqref{Lewis-inv} is the unique  invariant  which is quadratic in $q$ and $p$ by making a general ansatz with time dependent coefficients and imposing the Liouville invariance condition
\begin{equation}\label{Liouville-cond}
  \frac{d I}{dt}=\frac{\partial I}{\partial t}+\frac{1}{i\hbar}[I,H], \quad I^{\dag}=I.
\end{equation}
The authors took advantage of the eigenvalues (same as the time-independent harmonic oscillator) of this operator and its invariance property of taking solutions into solutions to study properties of TDHO. This approach requires some complicated arguments.  In a recent paper with pedagogical approach \cite{CoelhoQueirozAlves2025}, this method is reviewed with an application to the analysis of a quantum particle in a Paul trap. We also refer to \cite{Pedrosa1997, CoelhoQueirozAlves2022} for revisits of the TDHO.

The purpose of this article is to revisit the TDHO using group-theoretical methods. Within this framework, we easily and efficiently reobtain known results for the TDHO equation. Additionally, we discuss point transformation relations between the free Schrödinger and TIHO and TDHO equations in a unified manner using the most general transformations. In particular, we reconstruct the Schrödinger propagator for both the free particle and harmonic oscillator. This is done in Section 2. We devote Section 3 to an analysis of symmetries of the TDHO by two different  methods.  One is  Lie's prolongation process, the other one is by  commuting linear operators. Both methods are useful for deriving interesting invariant and separable solutions, and novel properties of the LR invariant in this context. Finally, we present a discussion of conservation laws for TDHO.

\section{Equivalence transformations}
Throughout this paper we will deal with the one-dimensional Schrödinger equation
\begin{equation}\label{oscillator}
  i \psi_t+\psi_{xx}=\omega(t)x^2 \psi
\end{equation}
with the corresponding Hamiltonian defined by
\begin{equation}\label{Ht}
 H(t)=p^2+\omega(t)x^2,
\end{equation}
where the
position and momentum operators, given by $x$ and $p=-i \partial_x$ satisfy
the canonical commutation relation $[x,p]=i$. So, Eq. \eqref{oscillator} can be expressed as
\begin{equation}\label{Schr}
   i \psi_t=H(t)\psi.
\end{equation}
Obviously, under an appropriate scaling, the frequencies of Hamiltonians \eqref{hamilt} and \eqref{Ht} are related by $\omega(t)=\Omega^2(t)/4$.

First of all, we look for equivalence transformations mapping \eqref{oscillator} into the same form, but changing only $\omega(t)$. To this end, we utilise a special case of the general result: Corollary 1 of   \cite{OezemirGuengoer2013} (See also \cite{GagnonWinternitz1993}) obtained for a nonlinear partial differential equation (PDE). Point transformations preserving the form of the linear Schrödinger equation with a complex potential $V(x,t)=V_1(x,t)+i V_2(x,t)$
\begin{equation}\label{gen-potential}
  i \psi_t+\psi_{xx}+V(x,t) \psi=0
\end{equation}
is given by  invertible point transformations   of the form
\begin{equation}\label{equiv-trans}
  \tilde{t}=T(t),  \quad \tilde{x}=X(x,t),  \quad \psi=Q(x,t)\tilde{\psi}, \quad Q=R_0(t)e^{i \theta(x,t)},
\end{equation}
where the transformed potential is related to the old one by
\begin{equation}\label{potential-relation}
 \tilde{V}(\tilde{x},\tilde{t})=\dot{T}^{-1}[V_1-\theta_t-\theta_x^2+i(V_2+\frac{\dot{R}_0}{R_0}+\theta_{xx})].
\end{equation}
Choosing
$$V_1(x,t)=-\omega(t)x^2, \quad \tilde{V}_1(\tilde{x},\tilde{t})=-\tilde{\omega}(\tilde{t})\tilde{x}^2, \quad  V_2(x,t)=\tilde{V}_2(\tilde{x},\tilde{t})=0$$
we find that
$X$ and $Q$ are further restricted
\begin{equation}\label{X-Q}
  X=\sqrt{\dot{T}}(x+\xi(t)),  \quad R_0(t)=\nu_0 \dot{T}^{1/4}, \quad \theta=-\frac{\ddot{T}}{8\dot{T}}x^2-\frac{\dot{\xi}}{2\sqrt{\dot{T}}}x+\eta(t),
\end{equation}
where $\dot{T}>0$, and
$\xi$ satisfies
\begin{equation}\label{xi}
  \ddot{\xi}+4\omega \xi=0.
\end{equation}

Consequently, the full transformation is given by
\begin{equation}\label{full-equiv}
 \psi(x,t)=\nu_0 e^{i \alpha} {\dot{T}^{1/4}} \exp{\left[i\left(-\frac{\ddot{T}}{8\dot{T}}(x+\xi)^2-\frac{\dot{\xi}}{2}x +\int(\omega \xi^2-\frac{1}{4}\dot{\xi}^2)dt\right)\right]}\tilde{\psi}(\tilde{x},\tilde{t}),
\end{equation}
where, by \eqref{potential-relation}, $\tilde{\omega}$ and $\omega$ are related by the relation
\begin{equation}\label{omega-tr}
  \tilde{\omega}(\tilde{t})=\dot{T}^{-2}\left(\omega(t)-\frac{1}{8}\curl{T;t}\right),
\end{equation}
where $\curl{T,t}$ is the Schwarzian derivative of $T$ with respect to $t$
\begin{equation}\label{schwarz}
  \curl{T,t}=\frac{\dddot{T}}{\dot{T}}-\frac{3}{2}\left(\frac{\ddot{T}}{\dot{T}}\right)^2.
\end{equation}
By a further change of time by
$\dot{T}=\sigma(t)^{-2}$ and due to the relation
$$\curl{T;t}= -2\frac{\ddot{\sigma}}{\sigma}$$ we can express \eqref{omega-tr} in a somewhat more convenient form
\begin{equation}\label{gen-EP}
 \ddot{\sigma}+4\omega(t) \sigma=4\tilde{\omega}(\tilde{t})\sigma^{-3}.
\end{equation}
In terms of $\sigma(t)$, up to a multiplicative constant, Eq. \eqref{full-equiv} can be expressed as
\begin{equation}\label{full-equiv-sigma}
   \psi(x,t)= {\sigma^{-1/2}} \exp{\left[i\left(\frac{\dot{\sigma}}{4\sigma}(x+\xi)^2-\frac{\dot{\xi}}{2}x +\int(\omega \xi^2-\frac{1}{4}\dot{\xi}^2)dt\right)\right]}\tilde{\psi}(\tilde{x},\tilde{t}),
\end{equation}
where
$$\tilde{x}={\sigma}^{-1}(x+\xi),  \quad \tilde{t}=\int_{t_0}^{t}\sigma(s)^{-2}ds.$$

We note that when $\tilde{\omega}(\tilde{t})=\omega(\tilde{t})$, the corresponding transformation is the full symmetry group of the oscillator equation.
For $\tilde{\omega}=\text{const.}$, this is the celebrated Ermakov--Pinney equation that can be solved in terms of two independent solutions of the associated  second order homogeneous  equation.

Some interesting implications of the relation \eqref{omega-tr} or \eqref{gen-EP}:

\begin{enumerate}[leftmargin=*]
  \item Case $\sigma=1$: Then for the choice  $\tilde{t}=t$, we have  $\tilde{\omega}(t)=\omega(t)$ and this leads to a subgroup of the full symmetry group of \eqref{oscillator}
\begin{equation}\label{sym-group-time-schr}
 \psi(x,t)=\nu_0 e^{i \alpha}  \exp{\left[i\left(-\frac{\dot{\xi}}{2}x +\int(\omega \xi^2-\frac{1}{4}\dot{\xi}^2)dt\right)\right]}\tilde{\psi}(x+\xi(t),t).
\end{equation}

 \item For the special choice $\tilde{\omega}=\omega=0$, we have  $\ddot{\sigma}=0$ or equivalently $\curl{T;t}=0$ and $\ddot{\xi}=0$. This implies $T(t)=M(t)$ is the Möbius transformations of the form
\begin{equation}\label{Moebius}
  M(t)=\frac{at+b}{ct+d}, \quad  ad-bc=1, \quad  \xi=vt+\beta.
\end{equation}
       So we have obtained the symmetry group of the free Schrödinger equation. This transformation formula is useful for obtaining the Green's function the free Schrödinger equation by mapping a constant solution of the free Schrödinger equation, for example $\tilde{\psi}(\tilde{x},\tilde{t})=1$ with $\xi=-y$ (constant).
Then we get
\begin{equation}\label{propaga-free}
 \psi(x,t)=C (ct+d)^{-1/2}\exp\curl{\frac{ic (x-y)^2}{4(ct+d)}}.
\end{equation}
If we compare this with the distributional limit
\begin{equation}\label{dirac}
  \delta(x)=\lim_{\varepsilon\to 0}\frac{1}{\sqrt{4\pi \varepsilon}}\exp\left[-\frac{x^2}{4\varepsilon}\right]
\end{equation}
we see that $C=(4\pi i)^{-1/2}$ and shifting $t=-d/c\to 0$  produces the fundamental solution of the free Schrödinger equation at the singularity $(x,t)=(y,0)$
\begin{equation}\label{Green}
  K(x,t;y)=(4\pi i t)^{-1/2}\exp\curl{\frac{i (x-y)^2}{4t}}.
\end{equation}

\item On the other hand for $\ddot{\sigma}=0$, $T(t)=M(t)$ as in \eqref{Moebius} we have the relation
$$(ct+d)^{-4}\tilde{\omega}(M(t))=\omega(t)$$
and choosing $\tilde{\omega}=\text{const}.$ we find that the most general form of $\omega(t)$ that can be transformed to the constant potential $\omega$ in the form
\begin{equation}\label{omega-2-const}
  \omega(t)=\frac{\omega_0^2}{(t+t_0)^4},  \quad \omega_0\ne 0.
\end{equation}
The parameter $t_0$ can be put to zero by a translation of time so that we have $\omega(t)=\omega_0^2 t^{-4}$.
In this case, $\xi$ is the solution to the classical TDHO equation (see \eqref{sol-basis} for its general solution)
\begin{equation}\label{clas-TDHO}
  \ddot{\xi}+4\omega_0^2 t^{-4}\xi=0.
\end{equation}
The simplest form of the relevant transformation  is given by
\begin{equation}\label{tr-const-freq}
  \psi(x,t)=t^{-1/2}\exp\left[\frac{i x^2}{4t}\right]\tilde{\psi}\left(\frac{x}{t},-\frac{1}t\right).
\end{equation}

\item More generally, we can use \eqref{gen-EP} to map the TDHO equation to the free Schrödinger equation by choosing $\tilde{\omega}=0$ or to the TIHO equation by choosing $\tilde{\omega}=1$. A special form of the second possibility where $\xi=0$  was used some time ago in \cite{Ciftja1999} to find the wave functions via an ansatz.

    In our approach, it is sufficient to put $\tilde{\omega}=\Omega^2=\text{const.}$, $\xi=0$ in \eqref{full-equiv-sigma} and \eqref{gen-EP} and to use the wave functions obtained by the separation of variables  in the form
\begin{equation}\label{wave-TIHO}
   \tilde{\psi}(x,t)=\tilde{\Psi}_n(x,t)=\frac{1}{\sqrt{2^n n!}}\left(\frac{\Omega}{\pi}\right)^{1/4}\exp\left[-i\Omega(2n+1) t-\frac{\Omega}{2}x^2\right]
 H_n(\sqrt{\Omega}x)
\end{equation}
corresponding to the time-independent  Hamiltonian $H=-\partial_x^2+\Omega^2 x^2$. Here $H_n(x)$, $n\in \mathbb{Z}^{\geq 0}$ are the Hermite polynomials. In this case, the normalized wave functions of the TDHO are given by
\begin{equation}\label{wave-TDHO}
   \Psi_n(x,t)= {\sigma^{-1/2}} \exp{\left[i\frac{\dot{\sigma}}{4\sigma}x^2\right]}\tilde{\Psi}_n\left(\frac{x}{\sigma},\int_{t_0}^{t}\sigma(s)^{-2}ds\right),
\end{equation}
where $\sigma$ is a solution to the EP equation
\begin{equation}\label{EP-sigma}
  \ddot{\sigma}+4\omega(t) \sigma=4\Omega^2\sigma^{-3}.
\end{equation}
More explicitly, the wave functions in (2.23) can be written in the form
\begin{align}\label{wave}
\Psi_n(x,t)
={}&
\frac{(\Omega/\pi)^{1/4}}{\sqrt{2^n n!\,\sigma(t)}}\,
H_n\!\left(\frac{\sqrt{\Omega}\,x}{\sigma(t)}\right)
\nonumber\\
&\times
\exp\left[
-\frac{\Omega x^2}{2\sigma(t)^2}
+i\frac{\dot{\sigma}(t)}{4\sigma(t)}x^2
-i\Omega(2n+1)\int_{t_0}^{t}\frac{ds}{\sigma(s)^2}
\right],
\end{align}
where $\Omega>0$ and $\sigma(t)>0$ satisfies \eqref{EP-sigma}.
For $\omega=\omega_0^2 t^{-4}$, a special solution of \eqref{EP-sigma} is $\sigma=t$ ($\Omega=\omega_0$), then the solution \eqref{wave-TDHO} agrees with \eqref{tr-const-freq} for $\tilde{\psi}=\tilde{\Psi}_n$.

These functions satisfy
\begin{equation}\label{Schr-PDE}
i\partial_t\Psi_n=H(t)\Psi_n.
\end{equation}
The constants $\Omega(2n+1)$ in \eqref{wave} are the eigenvalues
of the Lewis--Riesenfeld invariant
\begin{equation}\label{LR-inv}
I(t)=
\left(\sigma(t)p-\frac{\dot{\sigma}(t)}{2}x\right)^2
+\frac{\Omega^2}{\sigma(t)^2}x^2,
\qquad p=-i\partial_x.
\end{equation}
satisfying the eigenvalue problem
\begin{equation}
I(t)\Psi_n=\Omega(2n+1)\Psi_n,
\qquad
\partial_t I+i[H,I]=0.
\end{equation}
The values $\Omega(2n+1)$ can be interpreted as  the eigenvalues of the constant-frequency oscillator related to the TDHO by the transformation.
The time-dependent phase can be read off from \eqref{wave-TDHO} as
\begin{equation}\label{phase}
  \alpha_n(t)=-\Omega (2n+1)\int_{t_0}^{t}\sigma(s)^{-2}ds.
\end{equation}

If $\omega(t)=\omega_0^2>0$ is constant, choosing
$\sigma=\sqrt{\Omega/\omega_0}$ in \eqref{wave} yields
\begin{equation}
\Psi_n(x,t)=e^{-i(2n+1)\omega_0(t-t_0)}\phi_n(x),
\end{equation}
where
\begin{equation}\label{phi-n}
  \phi_n(x)=\omega_0^{1/4}\pi^{-1/4}(2^n n!)^{-1/2} \exp\left[-\frac{\omega_0 x^2}{2}\right]H_n(\sqrt{\omega_0}x),
\end{equation}
which recovers the usual stationary states.

The general solution of \eqref{EP-sigma} is expressible in the form
$$\sigma^2=A u_1^2+2B u_1u_2 +C u_2^2,  \quad \Omega^2=\frac{1}{4}(AC-B^2)W^{-2},$$ where $W$ is the constant Wronskian of the fundamental system of solutions $\curl{u_1,u_2}$ of the homogeneous ODE $\ddot{\sigma}+4\omega(t) \sigma=0$. See Ref. \cite{GuengoerTorres2017} for more details of this equation and its extensions.

We comment that the instantaneous eigenvalue problem
\begin{equation}
H(t)\phi_n(x,t)=E_n(t)\phi_n(x,t)
\end{equation}
has, at each fixed time, the eigenvalues
\begin{equation}
E_n(t)=\sqrt{\omega(t)}(2n+1), \quad \omega(t)>0
\qquad n=0,1,2,\ldots.
\end{equation}

\item Moreover, in \eqref{gen-EP}
putting $\omega(t)=\omega_0^2$, $\tilde{\omega}=0$ we obtain the transformation formula that maps the harmonic oscillator equation to the free Schrödinger equation. Here $T$ and $\xi$ satisfy
$$\curl{T;t}=8\omega_0^2,  \quad \ddot{\xi}+4\omega_0^2\xi=0.$$
Note that the general solution of the first equation (known as Schwarzian equation)
is expressible as ratios of two independent solutions of
$$\ddot{\sigma}+4\omega_0^2 \sigma=0.$$

Again by a proper choice of $\sigma$ and $\xi$ as $\sigma(0)=0$, $\dot{\sigma}(0)=2\omega_0$ and $\xi(0)=-y$, $\dot{\xi}(0)=0$ so that
$$\sigma(t)=\sin 2\omega_0 t, \quad  T(t)=-\frac{1}{2\omega_0}\cot 2\omega_0 t, \quad \xi=-y \cos 2\omega_0 t$$
we can map a constant solution of the free Schrödinger equation
by the formula \eqref{full-equiv-sigma} to the propagator of the oscillator equation, up to a multiplicative constant,
\begin{equation}\label{propaga}
  \psi(x,t)=C (\sin 2\omega_0 t)^{-1/2}\exp\curl{\frac{i \omega_0}{2}[(x^2+y^2)\cos 2\omega_0 t-2xy]\cosec 2\omega_0 t}
\end{equation}
This is exactly  Mehler's kernel $K(x,t;y)$ (also known as Schrödinger or Feynman propagator, Green's function) for the oscillator equation for the choice $C=\omega_0^{1/2}(2\pi i)^{-1/2}$ and satisfies the distributional limit
$$\lim_{t\to 0^{+}}K(x,t;y)=\delta(x-y).$$

A re-derivation of this kernel using group-invariant solutions will be presented in Section 3 (See formula \eqref{Mehler-inv}).

For alternative derivations of this kernel  we refer the interested reader to   References \cite{Beauregard1966, Holstein1998, BaroneBoschiFilhoFarina2003, LopezSuslovVegaGuzman2013}. For a similar derivation of  Mehler's (or heat)  kernel for the heat equation with variable potentials we refer to \cite{Guengoer2018, Guengoer2018a}. A discussion of (Schrödinger) propagators for the TDHO is given in \cite{VergelVillasenor2009}.

\item Finally, we put $\omega=0$ and $\tilde{\omega}=1$ in \eqref{gen-EP}, which reduces to
the simple EP equation $\ddot{\sigma}=4\sigma^{-3}$ with the general solution
$$\sigma^2=At^2+2Bt+C,  \quad (AC-B^2)W^{-2}=4,$$ where $W=-1$ is the Wronskian of $\curl{t,1}$. A convenient choice is $(A,B,C)=(1,0,4)$ so that $\sigma^2=1+4t^2$. Hence, by the transformation formula with $\xi=0$ we can conversely map from the free Schrödinger equation solution $\psi$ to the solution $\tilde{\psi}$ of the TIHO equation:
\begin{equation}\label{map-free-HO}
   \tilde{\psi}(\tilde{x},\tilde{t})= (1+4t^2)^{1/4}\exp\left[\frac{-i t x^2}{1+4t^2}\right]\psi(x,t),
\end{equation}
$$ x=(1+4t^2)^{1/2}\tilde{x}=\frac{\tilde{x}}{\cos 2\tilde{t}},  \quad t=\frac{1}{2}\tan (2\tilde{t}).$$
This transformation first appeared in \cite{Niederer1973} using a different argument. See also \cite{GagnonWinternitz1993}.
\end{enumerate}

We mention a recent paper \cite{AlcalaPlyushchay2026} that partly discusses similar transformation properties among variants of the Schrödinger equation.

\section{Symmetry algebra and invariant solutions}
In this section we calculate and identify Lie symmetry  algebra of \eqref{oscillator}. This can be realised by two different methods. The first one is to use the Lie prolongation formalism. A general element of symmetry algebra is taken in the form of vector fields (first order operators)
\begin{equation}\label{gen-vf}
  \mathbf{v}=\tau \gen t+\xi \gen x+\eta \gen \psi+\eta^* \gen {\psi^*},
\end{equation}
where the coefficients are all functions of $x,t,\psi,\psi^*$. Applying  Lie prolongation process we find that the finite part of the symmetry algebra $\lie$ is generated by
\begin{equation}\label{Lie-alg}
  \lie=\left< X(\tau(t)), Y(\chi(t)), M \right>,
\end{equation}
where
\begin{equation}\label{sym-vf}
  X(\tau(t))=\tau \gen t+\frac{\dot{\tau}}{2}x\gen x+\left[\frac{i\ddot{\tau}}{8}x^2-\frac{\dot{\tau}}{4}\right]\psi\gen {\psi}+\text{c.c.}
\end{equation}
and
\begin{equation}\label{heis}
  Y(\chi(t))=\chi\gen x+\frac{\dot{\chi}}{2}xM,  \quad M=i(\psi\gen \psi-\psi^{*}\gen {\psi^*}).
\end{equation}
A general element of the symmetry algebra can be written as
\begin{equation}\label{gen-element}
  \mathbf{v}= X(\tau(t))+Y(\chi(t))+aM.
\end{equation}
Here $\tau$ and $\chi$ satisfy a third order and second order homogeneous linear ODEs, respectively
\begin{equation}\label{3rd-ode}
  \dddot{\tau}+16\omega \dot{\tau}+8 \dot{\omega}\tau=0,  \quad \ddot{\chi}+4\omega(t)\chi=0.
\end{equation}
If $\omega(t)$ is left invariant under the transformation \eqref{full-equiv}, the first ODE is precisely the infinitesimal form of the condition \eqref{omega-tr}. This is seen by expanding $T$ and $\omega(\tilde{t})$ into the Taylor series near the group parameter $\varepsilon$
$$\tilde{t}=T(t;\varepsilon)=t+\varepsilon \tau+\mathcal{O}(\varepsilon^2),  \quad \tilde{\omega}(\tilde{t})=\omega(\tilde{t};\varepsilon)=\omega(t)+\varepsilon \tau \dot{\omega}+\mathcal{O}(\varepsilon^2) $$
and keeping terms linear in $\varepsilon$
$$\dot{T}^2\tilde{\omega}=\omega+[2\omega \dot{\tau}+\tau \dot{\omega}]\varepsilon+\mathcal{O}(\varepsilon^2)=\omega-\frac{1}{8}\varepsilon \dddot{\tau}+\mathcal{O}(\varepsilon^2).$$
This ODE enjoys many remarkable properties. Detailed discussion of them can be found in \cite{GuengoerTorres2017, CarinenaGuengoerTorres2020}.
One useful property is that it admits a basis of solutions $\{\phi_1^2,\phi_1\phi_2,\phi_2^2\}$ in terms of two independent solutions $\{\phi_1,\phi_2\}$ of the stationary Schrödinger equation
\begin{equation}\label{stat-schr}
  \ddot{\phi}+4\omega(t)\phi=0.
\end{equation}
Second equation of \eqref{3rd-ode}  corresponds to  $T=t$,  $\xi=\varepsilon \chi$ in infinitesimal form.

Another noteworthy fact is that it can be transformed to the Laguerre--Forsyth (LF) canonical form, where the LF invariant vanishes,  $\tilde{\tau}_{\tilde{t}\tilde{t}\tilde{t}}=0$ by an orientation-preserving change  of variables $(t,\tau)\to (\tilde{t},\tilde{\tau})$
\begin{equation}\label{change-variables}
  \tilde{t}=s(t),  \quad \tau(t)=\dot{s}(t)^{-1} \tilde{\tau}(s),
\end{equation}
where $s(t)$ satisfies the Schwarzian equation $\curl{s,t}=8\omega(t)$. However, for our purposes we prefer to keep the initial form \eqref{3rd-ode}.

Being $\chi_1,\chi_2$ linearly independent solutions of \eqref{stat-schr}, the generators $P=Y(\chi_1), B=Y(\chi_2), M$ span a Heisenberg algebra with commutators
$$[Y(\chi_1),Y(\chi_2)]=\frac{1}{2}W_\chi M,$$
where $W_\chi$ is the Wronskian of $\chi_1,\chi_2$. Commutation relations between $X(\tau(t))$ and $Y(\chi(t))$ are
$$[X(\tau(t)),Y(\chi(t))]=Y(\tau \dot{\chi}-\frac{1}{2}\chi \dot{\tau}).$$

The following trivial generators representing homogeneity and linear superposition should be added to the above algebra $\lie$ of \eqref{Lie-alg}
\begin{equation}\label{superpos}
  N=\psi\gen \psi+\psi^{*}\gen {\psi^*},  \quad Z(\varphi)=\varphi(x,t)\gen \psi,
\end{equation}
where $\varphi(x,t)$ is any solution to \eqref{oscillator}.
Apparently, in the free particle case $\omega=0$, symmetry algebra $\lie_0$ is the six-dimensional Schrödinger algebra spanned by the vector fields
\begin{equation}\label{Sch-alg}
  \lie_0=\left< X(1), X(t), X(t^2), Y(1), Y(t), M \right>.
\end{equation}
We mention that the other method for finding symmetries  is the formalism of commuting symmetry operators. In this method, we look for symmetry generators as differential operators in the form
\begin{equation}\label{sym-op}
  X=\tau(x,t)\gen t+\xi(x,t) \gen x+ f(x,t).
\end{equation}
Operators $X$ are required to commute with the linear operator
\begin{equation}\label{L-op}
  L=i\partial_t+\partial_x^2-\omega(t) x^2
\end{equation}
on the solution set of $L \psi=0$
\begin{equation}\label{comm-op}
  [L,X]=\lambda L,
\end{equation}
where $\lambda$ can be a constant,  a function of $x,t$, or a lower-order differential operator. $X$ will belong to the symmetry algebra. In the terminology of Miller, $X$ is called a conformal symmetry. If $\lambda\equiv 0$, $X$ is an ordinary symmetry.   By equating coefficients of $\partial_x^2, \partial_x, \partial_t, 1$   on both sides of \eqref{comm-op}, we obtain a system of differential equations for $\tau, \xi, f, \lambda$.

In our case, calculating the following commutators
\begin{equation}\label{two-comm}
 [L,X(\tau)]=\dot{\tau}L,  \quad [L,Y(\chi)]=0,
\end{equation}
we see that $\lambda=\dot{\tau}, 0$, respectively.

The Lie prolongation formalism is related to the formalism of commuting operators in a simple way:
\begin{equation}\label{vf-sym-rel}
  \eta(x,t,\psi,\psi^*)=-f(x,t)\psi-f^{*}(x,t)\psi^*.
\end{equation}
Further details of this relation can be found in \cite{LeviVinetWinternitz1997}.
The main disadvantage of this method is that it is only applicable to linear equations. Moreover, even for the linear equations the method may not give all symmetries. This is the method adopted in the excellent monograph \cite{Miller1984}. For example, the Schrödinger algebra $\lie_0$ in terms of symmetry operators are
\begin{equation}\label{free-schr-alg}
\begin{split}
   &  X_1=T=\gen t, \quad X_2=D=t\gen t+\frac{x}{2}\gen x+\frac{1}{4},  \quad X_3=C=t^2\gen t+xt\gen x-\left(\frac{i}{4}x^2-\frac{t}{2}\right), \\
    & X_4=P=\gen x, \quad X_5=B=t\gen x-\frac{ix}{2},  \quad X_6=M=i.
\end{split}
\end{equation}
It is straightforward to see that from \eqref{two-comm}   $\lambda=1$ for $X_2$, $\lambda=2t$ for $X_3$ and $\lambda=0$ in all other cases. Using notation of \cite{Miller1984}, we have
$$X_1=K_{-2}, \quad X_2=\frac{1}{2}K^{0}, \quad X_3=-K_2, \quad X_4=K_{-1},  \quad X_5=-K_1, \quad X_6=K_0. $$

One- and two-dimensional subalgebras of  the Schrödinger algebra $\lie_0$  using general methods developed by Winternitz and his collaborators under the action of the subgroup generated by the corresponding Lie algebra $\lie_0$ was performed in \cite{BoyerSharpWinternitz1976a}. Those subalgebras obtained  partition $\lie$ into orbits of linear operators. To each orbit there corresponds a coordinate system in which the Schrödinger equation admits solutions with separation.  A classification of all possible coordinate systems in which the free Schrödinger equation and its various extensions to nonzero potentials with sufficient nontrivial symmetry possess separable solutions was discussed in the same book.

For the sake of completeness, we give a list of the one- and two-dimensional subalgebras (See \cite{BoyerSharpWinternitz1976a} for the full classification):
\begin{multline}\label{one-dim-sub}
  \lr {T+C},  \quad \lr {T+C+a M}, \quad \lr {D},  \quad \lr {D+b M},  \quad \lr T, \quad \lr P,  \quad \lr M, \\
  \lr {T+\epsilon M},  \quad \lr{T+B},  \quad a\ne 0, \quad b>0, \quad \epsilon=\pm 1,
\end{multline}
\begin{multline}\label{two-dim-sub}
 \lr{T+C, M},  \quad  \lr{D, T}, \quad \lr{D, M},  \quad \lr{D, P}, \quad \lr{D+aM,T}, \quad \lr{T, M}, \\
  \lr{T, P} ,\quad \lr{P, M},  \quad \lr{D+aM, P}, \quad \lr{T+\epsilon M, P}, \quad \lr{T+B, M}.
\end{multline}

For the standard HO case when $\omega(t)=\omega_0^2$ (constant), Lie symmetry algebra is spanned by the vector fields
\begin{equation}\label{HO-algebra}
  \lie_{HO}=\lr {X(1),  X(\cos 4\omega_0 t),  X(\sin 4\omega_0 t),  Y(\cos 2\omega_0 t),  Y(\sin 2\omega_0 t),  M}.
\end{equation}

A more systematic approach to calculate Schrödinger propagator (Mehler's kernel) for the harmonic oscillator is the method of invariant solutions.
We look for our propagator $K(x,t;y)$ as a group invariant solution for the action $\mathbf{v}$ generated by a linear combination   of the elements of the five-dimensional subalgebra generated by $\lr {X(1),  X(\cos 4\omega_0 t),  X(\sin 4\omega_0 t), N, M}$
\begin{equation}\label{lin-comb}
  \mathbf{v}= k_1 X(1)+k_2  X(\cos 4\omega_0 t)+k_3  X(\sin 4\omega_0 t)+k_4 N+k_5 M,
\end{equation}
where $k_i, i=1,2,\ldots,5$ are a priori arbitrary constants. We require the initial condition  $K(x,0;y)=\delta(x-y)$ to be preserved by this symmetry. This condition restricts the $k_i$'s and hence the vector field $\mathbf{v}$ to be a one-dimensional symmetry
\begin{multline}\label{vf-bc}
  \mathbf{v}=2\sin^2 (2\omega_0 t) \gen t+2\omega_0\sin (4\omega_0 t)x\gen x+\\
  i\omega_0\left[-2\omega_0 y^2+2\omega_0 \cos (4\omega_0 t)x^2+i \sin (4\omega_0 t)\right]\psi\gen \psi.
\end{multline}
Invariant solution in terms of invariants $\zeta=(\sin 2\omega_0 t)^{-1}x$ and $F(\zeta)$ will have the form
\begin{equation}\label{Mehler-inv-form}
  \psi(x,t)= (\sin 2\omega_0 t)^{-1/2}\exp\left[\frac{i \omega_0 (x^2+y^2)}{2\tan (2\omega_0 t)}\right]F(\zeta).
\end{equation}
This solves the harmonic oscillator equation
$$L\psi=(i\partial_t+\partial_x^2-\omega_0^2 x^2)\psi=0$$
if $F(\zeta)$ satisfies the stationary harmonic oscillator equation
$$F''+\omega_0^2 y^2 F=0.$$
If we take the solution $F(\zeta)=C\exp{(-i \omega_0 y \zeta)}$, then the solution \eqref{Mehler-inv-form} turns out to be Mehler's kernel up to the multiplicative constant $C=[\omega_0/(2\pi i)]^{1/2}$ (See  \eqref{propaga})
\begin{equation}\label{Mehler-inv}
  \psi(x,t)=K(x,t;y)=\sqrt{\frac{\omega_0}{2\pi i \sin 2\omega_0 t}}\exp\left\{ \frac{i \omega_0 }{2\sin 2\omega_0 t}[(x^2+y^2)\cos 2\omega_0 t-2xy] \right\}.
\end{equation}
As one would expect, in the limit $\omega_0\to 0$, this kernel reduces to the free Schrödinger propagator  \eqref{Green}.

In the same spirit, the same subalgebra with direct-sum structure $\lie=\Sl(2,\mathbb{R})\oplus \lr {M, N}$ was successfully used to calculate   Mehler's kernel in terms of Bessel functions for a combined potential $V=kx^{-2}+\omega_0^2 x^2$ of harmonic oscillator and inverse-square potential \cite{Guengoer2021}.

We  observe that Eq. \eqref{3rd-ode} admits the first integral
\begin{equation}\label{first-int}
  K=\frac{1}{4}(2\tau \ddot{\tau}-\dot{\tau}^2)+4\omega \tau^2.
\end{equation}
The substitution $\tau(t)=\rho(t)^2$ in \eqref{first-int} converts it to the EP equation
\begin{equation}\label{EP}
\ddot{\rho}+4\omega(t) \rho=K \rho^{-3}.
\end{equation}
Then, for a given $\omega$ we obtain a three-dimensional symmetry algebra spanned by the vector fields
\begin{equation}\label{sl2}
  T=X(\tau_1)=X(\phi_1^2), \quad D=X(\tau_2)=X(\phi_1 \phi_2), \quad  C=X(\tau_3)=X(\phi_2^2)
\end{equation}
satisfying the $\Sl(2,\mathbb{R})$ commutation relations
\begin{equation}\label{comm}
  [T,D]=W T,  \quad [T,C]=2W D, \quad [D,C]=W C,
\end{equation}
where $W$ is the (constant) Wronskian of $\phi_1, \phi_2$. Of course, we can put $W\to 1$ by scaling of the three vector fields (This can be verified by direct calculation of the commutators).

In light of the above commutation relations we immediately identify the Lie symmetry algebra as a semidirect sum of the $\Sl(2,\mathbb{R})$ algebra generated by $T,D,C$ with the Heisenberg algebra  generated by $P,B,M$ as its ideal. The free Schrödinger equation and TDHO have isomorphic symmetry algebras.

A comprehensive group classification of (1+1)-dimensional linear Schrödinger equations with arbitrary complex potentials was performed in \cite{KurujyibwamiBasarabHorwathPopovych2018}.

The general element $X(\tau)$ of the symmetry algebra $\Sl(2,\mathbb{R})$ can be written as
\begin{equation}\label{gen-sl2}
  X(\tau)=a T+b D+c C.
\end{equation}
If we substitute
\begin{equation}\label{tau}
  \tau=\sigma^2=a \phi_1^2+b \phi_1 \phi_2+c \phi_2^2
\end{equation}
into the first integral  \eqref{first-int} we find the relation
\begin{equation}\label{K}
  K=-\frac{1}{4}\Delta W^2,  \qquad \Delta=b^2-4ac.
\end{equation}
$\Delta$ is an invariant of the full adjoint action of $\Sl(2,\mathbb{R})$.
One-dimensional subalgebras of the $\Sl(2,\mathbb{R})$ algebra depend on the sign of the invariant $\Delta$:
\begin{align}
        T &:   \quad \Delta=0 \quad (K=0) \\
        D &:  \quad \Delta>0 \quad (K<0) \label{K<0}\\
        T+C &:  \quad \Delta<0 \quad (K>0) \label{K>0}
\end{align}


Two interesting forms for the pairs $\omega, \tau$ leading to explicit solutions are
\begin{equation}\label{omega-1}
  \omega(t)=\frac{1}{4}\alpha(1-\alpha)t^{-2},  \quad \alpha\in \mathbb{R}\setminus \{0,1,1/2\},
\end{equation}
\begin{equation}\label{tau-1}
  \tau_1=\phi_1^2=t^{2(1-\alpha)}, \quad \tau_2=\phi_1 \phi_2=t,  \quad \tau_3=\phi_2^2=t^{2\alpha},  \quad W=2\alpha-1
\end{equation}
and
\begin{equation}\label{omega-2}
  \omega(t)=\omega_0^2 t^{\gamma},  \quad \gamma_{\pm}=\frac{-4k}{2k\pm 1}, \quad k\in \mathbb{N}.
\end{equation}
In the second case, general solutions of \eqref{stat-schr} can be expressed in terms of elementary functions.

For example, for the special frequency $\omega(t)=\omega_0^2 t^{-4}$ (see \eqref{omega-2-const}) we have the basis of solutions
\begin{equation}\label{sol-basis}
  \phi_1= t \cos(2\omega_0/t), \quad \phi_2=t \sin(2\omega_0/t), \quad W=-2\omega_0.
\end{equation}
We already know that this time-dependent frequency is equivalent to the constant frequency $\omega_0$ by the transformation \eqref{tr-const-freq}.
A  basis for the general solution of  $\tau$ is
\begin{equation}\label{tau-2}
  \tau_1=t^2\cos^2(2\omega_0/t),  \quad \tau_2=t^2\cos(2\omega_0/t)\sin(2\omega_0/t),  \quad \tau_3=t^2\sin^2(2\omega_0/t).
\end{equation}

The subalgebra  corresponding to $\tau=\tau_1+\tau_3=t^2$  is
$$T+C=X(\tau)$$ with $(a,b,c)=(1,0,1)$, $\Delta=-4<0$ ($K=4\omega_0^2>0$)
is especially of interest to us because it leads to $R$-separable solutions (Hermite $\times$ exponential). A solution is $R$-separable in coordinates $(u,v)$ if there is a function $R(x,t)$ such that \eqref{oscillator} admits solutions
$$\psi(x,t)=e^R \Psi=e^R \Psi_1(u)\Psi_2(v),$$
where $\Psi$ satisfies some eigenvalue equation \cite{Miller1984}.
According to Miller's method, the operator $X_0=\tau(x,t)\gen t+\xi(x,t) \gen x$ is transformed to semi-canonical form by choosing $u$ as an invariant satisfying $X_0(u)=0$, and leaving $v$ arbitrary so that $X_0=a(u,v)\gen v$.
In the above example the appropriate separable coordinates $(u,v)$ can be chosen as
\begin{equation}\label{coord}
  u=x \tau^{-1/2}=x (\phi_1^2+\phi_2^2)^{-1/2}=\frac{x}{t},  \quad v=t,  \quad X_0=\tau(v)\gen v.
\end{equation}
The corresponding separable solution is already included in \eqref{wave-TDHO} for $\sigma=t$, $\Omega=\omega_0$ with separation constant being $\lambda=\omega_0(2n+1)$.

Now we look into a situation in which the parameter in the  subgroup orbits appears as an eigenvalue in finding the group-invariant solution.
We let $K>0$ and choose the specific subalgebra $T+C+a M$ of the classification list \eqref{one-dim-sub}.
Two invariants of the vector field $X(\tau)+a M$, where $\tau$  is as given in \eqref{tau} with $K>0$  (See \eqref{K>0})  are found by solving the first order homogeneous linear PDE $(X(\tau)+a M)J(x,t,\psi)=0$ by the method of characteristics
\begin{equation}\label{invar}
  u=x \tau^{-1/2}, \quad \zeta=\psi \tau^{1/4} \exp\left[-\frac{i\dot{\tau}}{8\tau}x^2-ai \int_0^t \tau^{-1}(s)ds\right].
\end{equation}
We should look for invariant solutions in the form
\begin{equation}\label{inv-sol}
  \psi(x,t)=\tau^{-1/4} \exp\left[\frac{i\dot{\tau}}{8\tau}x^2+ai \int_0^t \tau^{-1}(s)ds\right]F(u).
\end{equation}
Substituting $\psi$ into \eqref{oscillator} we find that the invariant $F(\zeta)$ satisfies an ordinary differential equation
\begin{equation}\label{reduced-ode}
  F''(u)-\left\{a+\frac{1}{4}\left[\frac{1}{4}(2\tau \ddot{\tau}-\dot{\tau}^2)+4\omega \tau^2\right]u^2\right\}F(u)=0
\end{equation}
or using \eqref{first-int}
\begin{equation}\label{reduced-ode-2}
  F''-\left[a+\frac{K}{4}u^2\right]F=0.
\end{equation}
This ODE is known as Weber--Hermite (or parabolic cylinder) equation. We put $K=4\Omega^2$ ($\Delta=-16\Omega^2 W^{-2}<0$) for convenience. It is nothing more than the eigenfunction equation for the stationary quantum harmonic oscillator.  Its bounded (square-integrable) solutions can be expressed in terms of Hermite polynomials
\begin{equation}\label{hermite-sol}
  F(u)=F_n(u)=C_n \exp{\left(-\frac{\Omega}{2}u^2\right)}H_n(\sqrt{\Omega}u), \quad n=0,1,2,\ldots
\end{equation}
for the following values of $a$
$$a=-\Omega(2n+1).$$ We thus see that the fixed parameter $a$ figuring in the subalgebra plays the role of an eigenvalue. From the orthonormality property of Hermite polynomials we find $C_n=(\Omega/\pi)^{1/4}{(2^n n!})^{-1/2}$. Notice that the above procedure is exactly the same as solving the system $L\psi=0$, $X\psi=0$ simultaneously (Miller's approach) for the linear operator associated to $T+C+a M$. A crucial point that follows from this analysis is that the choice of one-dimensional subalgebra of the $\Sl(2,\mathbb{R})$ algebra determines the sign of $K$. This fact has been disregarded in all earlier occurrences of the eigenfunction formula \eqref{wave-TDHO} in the relevant literature.

In summary, the corresponding solution \eqref{inv-sol} with $F(u)$ given by \eqref{hermite-sol} agrees completely with the normalized wave functions \eqref{wave} for $\tau=\sigma^2$ as solution of \eqref{EP-sigma}.
This result clearly illustrates the importance of the subgroup classification.

We point out  the fact that the subalgebra $D+b M$ would lead to an ODE of the form \eqref{reduced-ode-2} with $K<0$ ($\Delta>0$) (See \eqref{K<0}). This case corresponds to an inverted harmonic oscillator (IHO). Unlike the standard HO, it has eigenstates that are not square-integrable (non-normalizable) on $\mathbb{R}$ though it has a continuous spectrum on the whole real line.  In spite of exhibiting a dynamics with quantum instability and chaos, it is exactly solvable in terms of Weber functions.

The case $K=0$ is trivial and corresponds to the free particle eigenvalue equation.

If we had taken the subalgebra $T+C$ (This means $a=0$ above), then  for arbitrary sign of $K\ne 0$, the change of variable $z=u^2$ would transform Eq. \eqref{reduced-ode-2} to
\begin{equation}\label{bessel-type}
  zF''(z)+\frac{1}{2}F'(z)-\frac{K}{16}z F(z)=0.
\end{equation}
The solutions of this ODE in terms of Bessel functions are given by
\begin{equation}\label{sol-bessel}
  F(z)=z^{1/4}Z_{\nu}(\alpha z),  \quad \nu=\pm \frac{1}{4},  \quad  \alpha^2=-\frac{K}{16},
\end{equation}
where $Z$ stands for the standard or modified Bessel functions depending on the sign of $\alpha^2$.  As $Z_{-1/4}$ is unbounded near zero we discard the solution for $\nu=-1/4$. Finally, we have constructed the exact solutions of \eqref{oscillator}  in the form
\begin{equation}\label{inv-sol-Bessel}
  \psi(x,t)=\sqrt{\frac{x}{\tau}}\exp\left[\frac{i\dot{\tau}}{8\tau}x^2\right]Z_{1/4}\left(\alpha \frac{x^2}{\tau} \right)
\end{equation}
or in terms of $\sigma$
\begin{equation}\label{inv-sol-Bessel-sigma}
  \psi(x,t)=\frac{\sqrt{x}}{\sigma}\exp\left[\frac{i\dot{\sigma}}{4\sigma}x^2\right]Z_{1/4}\left(\alpha \frac{x^2}{\sigma^2} \right).
\end{equation}

\begin{example}
An example of $K>0$:
Let us choose
$$\omega(t)=-\frac{1}{16t^2}(3+16 \mu^2 t^4), \quad \mu\ne 0. $$
then as a basis for the solution space  of \eqref{stat-schr} we can take
$$\phi_1=t^{-1/2}e^{\mu t^2},  \quad \phi_2=t^{-1/2}e^{-\mu t^2},  \quad W=-4\mu$$
so that for $(a,b,c)=(1/2,0,1/2)$ we have
$$\tau=\frac{1}{2}(\phi_1^2+\phi_2^2)=\frac{\cosh 2\mu t^2}{t}, \quad \Delta=-1<0,  \quad K=4\mu^2>0.$$
So we have $\alpha^2=-K/16$ ($\alpha=\pm i \mu/2$) and in the solution \eqref{inv-sol-Bessel} $Z_{1/4}$ is replaced by the modified Bessel function $I_{1/4}$.

\end{example}

Solutions invariant under the Heisenberg algebra (Galilean symmetries) have the form
\begin{equation}\label{galilean-inv}
  \psi(x,t)=C \chi^{-1/2}\exp\left[\frac{i\dot{\chi}}{4\chi}x^2\right].
\end{equation}
The choice $\chi=t$ ($\omega=0$) leads again to the singular solution of the free Schrödinger equation (compare with \eqref{Green}).

On the other hand, by integrating the vector field \eqref{heis} of Galilean invariance we obtain solution generating formula
\begin{equation}\label{galile-inv-sol}
  \psi_1(x,t)=\exp\curl{-\frac{i\dot{\chi}}{2}\left(\epsilon x+\frac{\epsilon^2}{2}\chi\right)}\tilde{\psi_1}(\tilde{x},\tilde{t}),  \quad \tilde{x}=x+\epsilon \chi(t), \quad \tilde{t}=t,
\end{equation}
where $\epsilon$ is the group parameter.
This formula transforms solutions into each other. Now if we combine
\eqref{full-equiv} for the choice $\xi=0$
\begin{equation}\label{xi-0}
  \psi_0(x,t)=(\sin 2t)^{-1/2}\exp\curl{\frac{i}{2}(\cot 2t) x^2}\psi_0(\tilde{x},\tilde{t})
\end{equation}
with this formula for ${\tilde{\psi_1}}(\tilde{x},\tilde{t})=1$, $\epsilon=-y$ and $\chi=\cos 2t$ we find
\begin{equation}\label{alter-kernel}
  \psi(x,t)=\psi_0(x-y \chi(t),t) \psi_1(x,t;y),
\end{equation}
where
$$\psi_1(x,t;y)=\exp\curl{\frac{-i\dot{\chi}}{2}\left(-x y+\frac{y^2}{2}\chi\right)}.$$
The above formula turns out to be exactly the kernel obtained in \eqref{propaga}.

\subsection{Lewis invariant revisited}

We now reveal some interesting results on the LR  invariant from symmetry point of view.
First, the LR invariant for \eqref{Ht} with $x$ scaled by $x\to x/\sqrt{2}$
$$I(t)=\frac{1}{2}\left(\rho^{-2} \hat{x}^2+(2\rho \hat{p}-\dot{\rho}\hat{x})^2\right).$$
as a second order linear differential operator in $\partial_x$ takes the form
\begin{equation}\label{LR-Op}
  I(t)=-2 \rho^2 \gen x^2+2i \rho \dot{\rho}x \gen x+i \rho \dot{\rho}+\frac{1}{2}\left(\dot{\rho}^2+\rho^{-2}\right)x^2
\end{equation}
Notice that for $K=\omega=1$, $\rho^2=1/2$ is a constant solution of \eqref{EP}  and $I(t)$ simply coincides with the Hamiltonian $I_0=H=-\psi_{xx}+x^2 \psi$ of the TIHO equation.

The commutation of the operator $L$ of \eqref{L-op} with $I$ as above
\begin{equation}\label{comm-L-I}
  [L,I]=-(\ddot{\rho}+4\omega(t) \rho- \rho^{-3})[2\rho x \partial_x+(\rho-i \dot{\rho}x^2)]
\end{equation}
vanishes identically (namely $\lambda=0$ in \eqref{comm-op})
whenever $\rho$ satisfies the EP equation \eqref{EP} exactly when $K=1>0$ (this corresponds to a subalgebra where $\Delta<0$ in \eqref{K}) as to be expected. So the LR invariant  operator is a second order (ordinary) symmetry in the sense of commuting linear operators. If $\psi$ is a solution, then $I(t)\psi$ is again a solution. Of course, $I$ can be redefined by the replacement $\sigma\to K^{1/4}\sigma$ to include $K$ as an arbitrary parameter  so that \eqref{EP} is satisfied. From a different point of view we have confirmed the well-known invariance of the LR operator as noted in \cite{Lewis1968, LewisRiesenfeld1969}.

On the other hand, the action of $I\psi$ on the $\Sl(2,\mathbb{R})$-invariant solution \eqref{inv-sol} satisfies the relation
\begin{equation}\label{LR-expl}
  X(\sigma^2)(I\psi)=R_1[zF(z)-8 F'(z)-16zF''(z)]=R_1(1-K)z F(z),
\end{equation}
which vanishes exactly when $K=1$.
Here  we have used the Bessel-type invariant ODE in $F$ \eqref{bessel-type} and
\begin{equation}\label{R1}
  R_1(x,t)=\frac{1}{4}\sigma^{-1/2}\exp\left[\frac{i \dot{\sigma}}{4\sigma}x^2\right].
\end{equation}
Since $I(t)$ contains no time derivative operator we can set $a=0$ with no loss of generality in \eqref{LR-expl}.
This observation shows that on the particular group-invariant family con\-si\-der\-ed here, the Lewis–Riesenfeld operator defines a compatible second-order differential constraint. For the normalization K = 1 (invariance under a one-dimensional subalgebra $T+C$ of $\Sl(2,\mathbb{R})$ (the case \eqref{K>0})), the action of the
corresponding symmetry generator on this constraint vanishes along the invariant family.

Another notable result is arrived at by acting on $I\psi$ with $\psi$ being the same invariant solution \eqref{inv-sol} by the second order prolongation of the vector field \eqref{sym-vf}
\begin{equation}\label{2ndprolong}
   {\pr{2}}X(\sigma^2)(I\psi)=R_2[zF(z)-8 F'(z)-16zF''(z)]=R_2(1-K)z F(z),
\end{equation}
with $R_2$ being a function of $x,t$
\begin{equation}\label{R2}
  R_2(x,t)=\frac{1}{16}\sigma^{-1/2}\exp\left[\frac{i \dot{\sigma}}{4\sigma}x^2\right][(\ddot{\sigma}+\dot{\sigma}^2)x^2+2i\dot{\sigma}\sigma],  \quad \dot{\sigma}\ne 0.
\end{equation}

The relation \eqref{2ndprolong} again vanishes for $K=1>0$. This observation can be interpreted as an aspect of the LR operator as a second order differential invariant, namely the prolonged action of the symmetry vector field on $I\psi$ vanishes only on the invariant solution rather than on any function $\psi$.

\subsection{Variational symmetries and conservation laws}

We can use Noether's theorem to derive conservation laws once we have the symmetry algebra of the TDHO.  Eq. \eqref{oscillator} admits a Lagrangian
\begin{equation}\label{Lagran}
  L=\frac{i}{2}(\psi^* \psi_t-\psi \psi^*_t)-|\psi_x|^2-\omega(t) x^2 |\psi|^2.
\end{equation}
In this case,  energy is no longer conserved. But, still, other conservation laws exist.

We find it more convenient to work in the modulus-phase representation $$\psi=R(x,t)e^{i\phi(x,t)}. $$
In terms of $R,\phi$, Eq. \eqref{oscillator}  can be rewritten as a system of two real second order PDEs
\begin{equation}\label{Oscillator-mod-phase}
  \begin{split}
     & R_{xx}-R \phi_x^2-R \phi_t-\omega x^2 R=0, \\
      & R_t+2R_x \phi_x+R \phi_{xx}=0.
  \end{split}
\end{equation}
In turn, the first order Lagrangian $L$ (up to a multiplicative numerical factor) has the form
\begin{equation}\label{L-mod-phase}
  L=R_x^2+R^2 (\phi_x^2+\phi_t+\omega x^2).
\end{equation}
We now check if the symmetry generators \eqref{sym-vf} and \eqref{heis} are also variational. This is done by the variational invariance condition for the vector field \eqref{gen-vf}
\begin{equation}\label{var-sym-cond}
  {\pr{1}}X(\tau)(L)+\Div(\tau, \xi)L=0,
\end{equation}
where $\Div(\tau, \xi)=\tau_t+\xi_x$.
Generators $X(\tau)$, $Y(\chi)$, $M$ in the coordinates $\curl{R, \phi}$ are
\begin{equation}\label{generators-mod-phase}
  \begin{split}
     &  X(\tau)=\tau \gen t+\frac{\dot{\tau}}{2}x\gen x-\frac{\dot{\tau}}{4}R\gen R+\frac{i\ddot{\tau}}{8}x^2\gen {\phi},\\
      &  Y(\chi)=\chi\gen x+\frac{\dot{\chi}}{2}xM,  \quad M=\gen \phi.
  \end{split}
\end{equation}
We see that the condition \eqref{var-sym-cond} is satisfied for the algebra \eqref{Lie-alg} generated by $X(\tau)$, $Y(\chi)$ and $M$
$${\pr{1}}X(\tau)(L)+\frac{3\dot{\tau}}{2}L=\frac{1}{8}x^2 R^2(\dddot{\tau}+16\omega \dot{\tau}+8 \dot{\omega}\tau)\equiv 0,$$
$${\pr{1}}X(\chi)(L)=\frac{x}{2}R^2 (\ddot{\chi}+4\omega\chi)\equiv 0,$$
owing to \eqref{3rd-ode}.
So our symmetry algebra is also variational and by  Noether's theorem we can calculate conserved densities of  conservation laws in the  form
\begin{equation}\label{cons-dens}
  \mathcal{I}_{t}=\tau L+Q_R \frac{\partial L}{\partial R_t}+Q_\phi \frac{\partial L}{\partial \phi_t},
\end{equation}
and the fluxes
\begin{equation}\label{flux}
  \mathcal{I}_{x}=\xi L+Q_R \frac{\partial L}{\partial R_x}+Q_\phi \frac{\partial L}{\partial \phi_x},
\end{equation}
where $Q_R$, $Q_\phi$ are the characteristic functions of $X(\tau)$, $Y(\chi)$ defined by
$$Q_R=-\left(\frac{\dot{\tau}}{4}R+\tau R_t+\frac{x}{2}\dot{\tau}R_x\right),  \quad Q_\phi=\frac{\ddot{\tau}}{8}x^2-\tau\phi_t-\frac{\dot{\tau}}{2}x\phi_x,$$
for $X(\tau)$ and
$$Q_R=-\chi R_x, \quad Q_\phi=\frac{\dot{\chi}}{2}x-\chi \phi_x$$
for $Y(\chi)$.
On the solution of the system \eqref{Oscillator-mod-phase}, the divergence of the vector $(\mathcal{I}_{t},\mathcal{I}_{x})$ vanishes identically
$$\Div(\mathcal{I}_{t},\mathcal{I}_{x})=D_t \mathcal{I}_{t}+D_x \mathcal{I}_{x}\equiv 0.$$

According to formula \eqref{cons-dens} we find the conserved densities
\begin{equation}\label{cons-dens-Xtau}
 \begin{split}
     \mathcal{I}_t(\tau)&=\tau (R_x^2+R^2 \phi_x^2)+\frac{1}{8}(\ddot{\tau}+8\omega \tau)x^2 R^2-\frac{\dot{\tau}}{2}x R^2 \phi_x \\
     &=\tau |\psi_x|^2+\frac{1}{8}(\ddot{\tau}+8\omega \tau)x^2 |\psi|^2-\frac{\dot{\tau}}{2}x \Im(\psi_x \psi^*),
 \end{split}
\end{equation}
\begin{equation}\label{cons-dens-Ychi}
  \mathcal{\hat{I}}_t(\chi)=R^2(\frac{\dot{\chi}}{2}x -\chi \phi_x)=\frac{\dot{\chi}}{2}x |\psi|^2-\frac{i}{2}\chi (\psi {\psi}_x^*-\psi^* \psi_x)=\frac{\dot{\chi}}{2}x |\psi|^2-\chi \Im(\psi_x \psi^*).
\end{equation}
In the case of free Schrödinger equation $\omega=0$,  our Lie algebra is the Schrödinger algebra then we obtain the conserved quantities
\begin{equation}\label{free-TDC}
  \mathcal{E}=|\psi_x|^2,  \quad \mathcal{D}=t \mathcal{E}-\frac{x}{2}\mathcal{P},  \quad \mathcal{C}=t^2 \mathcal{E}-xt \mathcal{P}+\frac{1}{4}x^2 \mathcal{M},
\end{equation}
\begin{equation}\label{free-PBM}
  \mathcal{P}=\Im(\psi_x \psi^*), \quad \mathcal{B}=\frac{x}{2}\mathcal{M}-t \mathcal{P},  \quad \mathcal{M}=|\psi|^2.
\end{equation}
The quantities $\mathcal{E}, \mathcal{P}, \mathcal{B}, \mathcal{M}$ are associated to  conservation of energy, linear momentum, center of mass and mass (norm or probability density), respectively. The conserved charges are  their integrals over the entire line with respect to $x$. For example, conserved energy is given by the integral
$$E=\int_{\mathbb{R}}\mathcal{E}\;dx.$$

Note that the second equation of the system \eqref{Oscillator-mod-phase} is already in a form of conservation law. In order to see this, we multiply it by $2R$ and write
\begin{equation}\label{2nd-in-cons}
 \gen t R^2+2D_x(R^2 \phi_x)=0,
\end{equation}
where $\varrho(x,t)=R^2=|\psi|^2$ is the probability  density and
$$J(x,t)=2R^2 \phi_x=i(\psi {\psi}_x^*-\psi^* \psi_x)$$
is the probability current (flux). This is the conservation law corresponding to the invariance under the constant change of phase $\gen \phi$. The trivial scaling symmetry $N$ can not leave the action integral invariant because ${\pr{1}}(N)(L)=2L\ne 0$.

The above equation is recognised  as the continuity equation
\begin{equation}\label{continuity}
 \gen t \varrho+\gen x J=0.
\end{equation}
This equation does not depend on the form of potential in general.
Under the normalizability condition of the wave function $\psi$ and vanishing of $J(x,t)$ at infinity this equation gives the conservation  of total probability.

In a similar manner, for the standard harmonic oscillator $\omega=1$, five conserved quantities  can be obtained from \eqref{cons-dens-Xtau}-\eqref{cons-dens-Ychi} by the choice $\tau\in\curl{1,\cos 4t,\sin 4t}$ and $\chi\in\curl{\cos 2t,\sin 2t}$. In the case when $\tau=1$
$$\mathcal{I}_t(1)=R_x^2+R^2 (\phi_x^2+x^2)=|\psi_x|^2+x^2|\psi|^2$$
is the conserved energy.

We comment that  in  \cite{Lutzky1978} a similar approach was followed for the classical TDHO to derive the Lewis invariant \eqref{Lewis-inv} in the classical case. But, in the quantum case, this approach does not provide the invariant because it is of second order in derivatives and the Lagrangian is first order and  Noether's theorem provides only first order ones.

\section{Conclusions}
The main contributions of this study can be summarised as follows:
\begin{enumerate}[leftmargin=*]
\item The transformations among the TDHO, the constant-frequency harmonic oscillator, and the free
Schrödinger equation are presented systematically.

\item  Mehler’s kernel (Schrödinger propagator)  is derived both through point transformations and through the invariant solution method.

\item Lie symmetry algebras of TDHO are discussed using both Lie prolongation procedure and classical method of linear operators.

\item The use of the relationship among the sign of $K$, the  invariant $\Delta$ (See \eqref{K}), and  the one-dimensional subalgebra orbits  of $\Sl(2, \mathbb{R})$  is shown to play an important role in the study of TDHO. An appropriate subalgebra  determines the normalizable HO, inverted-HO, and parabolic/Bessel cases.

\item  The cases $K > 0$, $K < 0$, and $K = 0$ can be associated  with the standard oscillator, inverted oscillator, and free/parabolic cases. This aspect provides a useful classification perspective.

\item A new compatibility interpretation between the LR operator and the Lie-invariant
solution manifold is provided.

\item Conservation laws using Noether theorem  are obtained in modulus-phase variables.

\end{enumerate}

\section*{Acknowledgments}
The writing of this article was started while the first author (F.G.) was visiting Departamento de Física Teórica, Universidad de Valladolid. He thanks the Department  for their kind hospitality during his six-week stay in the summer of 2026.

\medskip

\noindent{\bf Conflict of interest statement}
\smallskip

\noindent The authors declares that there is no conflict of interest.

\medskip

\noindent{\bf Data availability statement}
\smallskip

\noindent The data that support the findings of this work are available within the article.


%
%
%
%

\end{document}